\documentclass[conference]{IEEEtran}
\IEEEoverridecommandlockouts
\usepackage{cite}
\usepackage{amsmath,amssymb,amsfonts}
\usepackage{amssymb}
\usepackage{algorithmic}
\usepackage{algorithm}
\usepackage{graphicx}
\usepackage{textcomp}
\usepackage{xcolor}
\usepackage{amsthm}
\usepackage{subfigure}
\usepackage{caption}
\usepackage{xcolor}
\usepackage{color}
\def\BibTeX{{\rm B\kern-.05em{\sc i\kern-.025em b}\kern-.08em
    T\kern-.1667em\lower.7ex\hbox{E}\kern-.125emX}}
\usepackage[numbers,sort&compress]{natbib}
\usepackage{hyperref}
\usepackage{enumerate}
\begin{document}

\title{Understanding the Gain of Deploying IRSs in Large-scale Heterogeneous Cellular Networks
}

\author{\IEEEauthorblockN{Hu Cheng$^{\dagger,\S}$, Hongguang Sun$^{\dagger,\S}$, Linyi Zhang$^{\dagger,\S}$, Jiahui Li$^{\dagger,\S}$, Xijun Wang$^{\curlyvee}$, and Tony Q. S. Quek$^{\ddagger}$}
\IEEEauthorblockA{$^{\dagger}$College of Information Engineering, Northwest AF University, Yangling, Shaanxi, China \\
$^{\curlyvee}$School of Electronics and Information Technology, Sun Yat-sen University, Guangzhou, Guangdong, China\\
$^{\ddagger}$Information System Technology and Design Pillar, Singapore University of Technology and Design, Singapore\\
$^{\S}$Key Laboratory of Agricultural Internet of Things, Ministry of Agriculture and Rural Affairs, Yangling, Shaanxi, China\\
E-mail: hgsun@nwafu.edu.cn, cheng\underline{ }gong\underline{ }mi@163.com, LinyiZhang2020@163.com, lijiahui\underline{ }qqmail@qq.com, \\xjwang22@gmail.com, tonyquek@sutd.edu.sg}
}

\maketitle

\begin{abstract}
As the superior improvement on wireless network coverage, spectrum efficiency and energy efficiency, Intelligent reflecting  surface (IRS) has received more and more attention. In this work, we consider a large-scale IRS-assisted heterogeneous cellular network (HCN) consisting of $K$ ($K \geq 2$) tiers of  base stations (BSs) and one tier of  passive IRSs. With tools from stochastic geometry, we analyze the coverage probability and network spatial throughput of the downlink IRS-assisted $K$-tier HCN.  Compared with the conventional HCN, we observe the significant gain achieved by IRSs in  coverage probability and network spatial throughput. 
 The proposed analytical framework can be used to understand the limit of gain achieved by IRSs in HCN.
\end{abstract}


\section{Introduction}
Recently, with the emergence of the concept of smart and reconfigurable wireless  environment \cite{ref6,ref7}, intelligent reflecting  surfaces (IRSs) are expected to become a promisingly reliable and cost-effective solution in the future wireless networks. 
 Deploying IRSs between a transmitter and a receiver, and properly adjusting the reflecting elements, the signal propagation can be reconfigured to realize the expected scenarios, such as passive beamforming \cite{ref1}. As a passive transmission plane, IRS has lower deployment cost and energy consumption than an active BS and the radio frequency (RF) chain. In addition, it has the characteristics of light weight and flexible shape combination \cite{ref8}. Therefore, IRSs are suitable for intensive deployment in wireless network environment to increase the coverage and capacity of the network at lower cost.\\
\indent For IRS-assisted wireless networks, the authors in \cite{ref10} investigated the IRS passive beamforming schemes. The authors in \cite{8989805} considered a single-cell scenario with multiple IRSs, and characterized spatial throughput. The authors in \cite{ref11} considered a single link scenario and analyzed the performance of the IRS-assisted point-to-point non-direct link.  However, the works mentioned above considered the analysis and optimization only from the link level or in small-scale network scenarios. For large-scale IRS-assisted wireless networks, given the locations of IRSs/BSs, the authors  in \cite{ref13} investigated the IRS quasi-static phase shift design based on channel state information (CSI).
In \cite{9110835}, the authors evaluated the performance of a millimeter wave (mmWave) network in which the average achievable rate was obtained by deriving the Laplace Transform of of the aggregated interference from all BSs and IRSs. \textcolor{black}{In \cite{10036459}, the authors considered an IRS-assisted cellular-based RF-powered Internet of Things (IoT) network to quantify the gain achieved by IRSs.} The authors in \cite{ref14, 9673721} proposed analytical frameworks to evaluate the performance of DL IRS-assisted cellular network in which the coverage probability, network spatial througput or ergodic capacity, and energy efficiency were derived. However, the works \cite{ref13,9110835,10036459,ref14,9673721} only considered a single-tier cellular network, and the proposed analytical results can not be applicable to the heterogeneous cellular network (HCN) architecture. As HCN is an essential network paradigm even in beyond 5G, it is of great importance  to understand the limit of gain achieved by IRSs in a HCN.\\
\indent Motivated by the abovementioned, in this work, we consider an IRS-assisted large-scale $K$-tier HCN scenario consisting of multiple tiers of BSs and one tier of passive IRSs. We propose a general analytical framework to assess the performance of an IRS-assisted hybrid downlink HCN, which allows to evaluate the gain achieved by IRSs with regards to coverage probability and network spatial throughput. Compared with the conventional HCN, the proposed framework reveals that adding IRSs can significantly improve the desired signal power, as well as slightly increase inter-tier and intra-tier interference power. As a result, substantial performance gain is achieved by IRSs, which grows with the densification of IRSs and tends to be saturated eventually.  

\section{System Model}
\subsection{Network Model} \label{subsec 2-0}
\begin{figure}[t]
	\vspace{-2em}
	\centering
	\includegraphics[height=2in,width=3.5in]{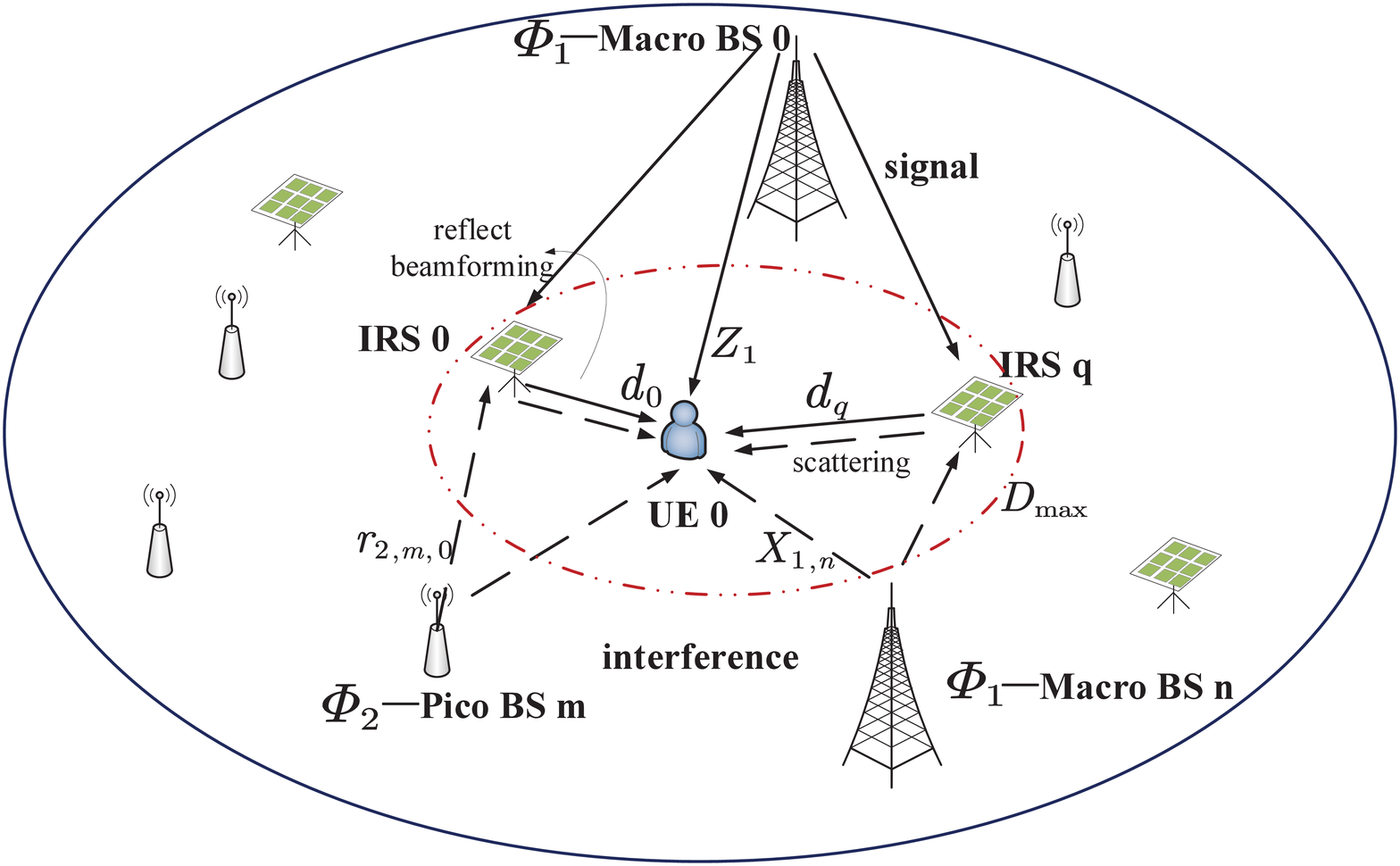}
	\caption{An illustration of an IRS-assisted two-tier macro/pico HCN in the  downlink.}
	\label{fi1}
	\vspace{-1em}
\end{figure}
As is shown in Fig. \ref{fi1}, we consider a downlink (DL) hybrid HCN comprising $K$ tiers of BSs, overlaid with one-tier passive IRSs. BSs of tier-$j$ $(j \in \{1, \dots , K\})$ have the same transmit power $P_{j}$ and height $H_{j}$, whose horizontal locations are modeled by a 2-dimensional (2D) homogeneous poisson point process (HPPP) $\Phi_{j}$ of spatial density $\lambda_{j}$. The height of IRS (with $N$ reflecting elements) is denoted by $H_{I}$, the horizontal locations of IRSs are scattered according to a 2D HPPP  $\Phi_{I}$ of spatial density $\lambda_{I}$.  
\textcolor{black}{BSs are assumed to acquire perfect CSI of all relevant channels. A time-division duplexing protocol is employed by the passive IRSs with the hypothesis of channel reciprocity for CSI acquisition.}
We adopt orthogonal multiple access technology, such that there is only one UE within a cell can be active at any given timeslot and subchannel.
 To study the effect of network traffic load, we  assume that BSs of the $j$-th tier has the same load factor $p_{j}$ $(0 < p_{j} \leq 1)$. As a result, the active BSs in the $j$-th tier that transmit on the same timeslot and subchannel can be modeled by a  thinned HPPP  $\Phi_{j}'$ with density $\lambda_{j}'=p_{j}\lambda_{j}$ .  \\
\indent According to Slivnyak's theorem, it is sufficient to focus on a \emph{typical} UE $0$ located at the origin who is assumed to be associated with BS $0$ in the $k$-th tier  ($ k \in \{1,\dots,K\}$). We define $X_{k,i}$, $d_{q}$ and $r_{j,m,q}$ as the 3D distance from the $k$-th tier BS $i$ to UE $0$, the 2D distance from the IRS $q \in \Phi_{I}$ to UE $0$, and the 2D distance from the BS $m \in \Phi_{j}$ to IRS $q$, respectively. Considering the limited reflective capability of a given IRS, we consider a practical local region of radius $D_{\mathrm{max}}$\footnote {The radius threshold $D_{\text{max}}$ is set based on the real situation so that each IRS can only provide services for a limited number of UEs nearby.} with the \emph{typical} UE $0$ being the center.  We denote IRSs within the local region by the set $\varDelta \triangleq \{q \in \Phi_{I}| d_{q} \leq D_{\mathrm{max}}\}$.  
If $\varDelta = \emptyset$, the \emph{typical} UE $0$ will be only served by its own BS.
We assume that only IRSs within the set $\varDelta$ can contribute interference to the \emph{typical} UE $0$.
\subsection{Cell Association Policy} \label{subsec 2-1}
We assume open access, and the association of a UE with a given tier is based on the maximum biased received signal power averaged over fading \cite{ref9}. We denote $B_{j}$ by the bias factor of the $j$-th tier which is adopted to balance the traffic load among BSs of different tiers. A \emph{typical} UE $0$ is associated with the nearest BS of tier $k$ if
\begin{equation}
	i= \mathrm{arg} \underset{k \in \{1,\dots,K\}}{\mathrm{max}} P_{k}B_{k} Z_{k}^{-\alpha_{k}},
	\label{formula11}
\end{equation}
where $Z_{k}$, $k \in \{1,\dots,K\}$ denotes the 3D distance from the \emph{typical} UE $0$ to the associated BS $0 \in \Phi_{k}$.
  Then, the probability that the \emph{typical} UE $0$ is associated with the $k$-th tier BS $0$ is given by \\
  \textcolor{black}{
\begin{equation}
	\small
	\begin{aligned}
		\mathcal{A}_{k} = 2\pi\lambda_{k} \int_{H_{k}}^{\infty} x\exp\{-\pi\sum_{j=1}^{K} \lambda_{j}[(\hat{P}_{j}\hat{B}_{j})^{\frac{2}{\alpha_{j}}}x^{\frac{2}{\hat{\alpha}_{j}}} - H_{j}^{2}]^{+}\} \mathrm{d}x,
		\label{formula13}
	\end{aligned}	
\end{equation}
}
 where $\hat{P}_{j} \triangleq P_{j}/P_{k}$, $\hat{B}_{j} \triangleq B_{j}/B_{k}$, $\hat{\alpha}_{j} \triangleq \alpha_{j}/\alpha_{k}$, and $[\cdot]^{+} \triangleq \mathrm{max}\{\cdot,0\}$. A modest modification of Lemma 1 in \cite{ref9} can be used to show the conclusion in (\ref{formula13}). Since the BSs of each tier are deployed independently according to a 2D HPPP, the probability density function (PDF) of $X_{k,0}$ is derived by
 \begin{equation}
 	\small
 	 \begin{aligned}
 		f_{X_{k,0}}(x)= \dfrac{2\pi\lambda_{k}}{\mathcal{A}_{k}}x\exp\Big\{-\pi  \sum_{j=1}^{K}\lambda_{j}[(\hat{P}_{j}\hat{B}_{j})^{\frac{2}{\alpha_{j}}}x^{\frac{2}{\hat{\alpha}_{j}}} - H_{j}^{2}]^{+}\Big\}.
 		\label{formula14}
 	\end{aligned} 
\end{equation}
\indent Moreover, under the condition that set $\varDelta$ is non-empty, we consider the policy that the \emph{typical} UE $0$ is associated with the nearest IRS $0$ within $\varDelta$ for dedicated reflect beamforming, such that the  PDF of the 2D distance $d_{0}$ from the \emph{typical} UE $0$ to its serving IRS $0$ is $f_{d_{0}}(d) = 2\pi\lambda_{I} d \mathrm{exp}\{-\pi\lambda_{I}d^{2}\}$.
\subsection{Channel Model} \label{subsec 2-2}
The channel model consists of large-scale fading and small-scale fading. For large-scale fading, the path loss exponents of BSs and IRSs are denoted by  $\{\alpha_{j} > 2\}_{j=1,\dots,K}  $ and $\alpha_{I} > 2 $, respectively. Small-scale fading is modeled by Rayleigh fading with unit power, such that the channel power gain follows an exponential distribution  denoted by $\eta \sim \mathrm{exp}(1)$. Define $\sigma^{2}$ as the additive white Gaussian noise (AWGN).\\
\indent For  BS $m$ in the $j$-th tier, the baseband equivalent channels from BS $m$ to UE $0$, from BS $m$ to IRS $q$ and from IRS $q$ to UE $0$ are denoted by $f_{\mathbf{d},m}^{(j)} \in \mathbb{C}$, $\mathbf{f}_{\mathbf{i},m}^{(j,q)} \in \mathbb{C}^{N \times 1}$ and $\mathbf{f}_{\mathbf{r}}^{(q)} \in \mathbb{C}^{1 \times N}$, respectively. We define  $\mathbf{\psi}^{(q)} \triangleq [\psi_{1}^{(q)},\dots, \psi_{N}^{(q)}]$ and define a diagonal matrix $\mathbf{\Psi}^{(q)} \triangleq \mathrm{diag}\{[e^{\mathbf{i}\psi_{1}^{(q)}},\dots,e^{\mathbf{i}\psi_{N}^{(q)}}]\}$ ($\mathbf{i}$ denotes the imaginary unit) as the phase shift coefficients matrix of the IRS $q$, where $\psi_{n}^{(q)} \in [0,2\pi)$ represents the phase shift of the reflecting element $n$. To maximize the  beamforming gain, we assume that the reflection coefficient of each reflecting element can achieve the unit amplitude \cite{ref10}. \\
\indent The \emph{typical} UE $0$ is related to two types of wireless links, i.e., the direct link BS-UE $0$ and the cascaded link BS-IRS-UE $0$.  According to the small-scale Rayleigh fading, we derive that $f_{\mathbf{d},m}^{(j)}$ follows the circularly symmetric complex Gaussian (CSCG) distribution, i.e., $\mathcal{CN} (0,l_{\mathbf{d},m}^{(j)})$, and the corresponding  channel power gain can be derived as 
\begin{equation}
	\vert f_{\mathbf{d},m}^{(j)} \vert^{2} \triangleq  \eta_{\mathbf{d},m}^{(j)} l_{\mathbf{d},m}^{(j)} = \eta_{\mathbf{d},m}^{(j)}\beta X_{j,m}^{-\alpha_{j}}, \label{formula1}
\end{equation}
where $l_{\mathbf{d},m}^{(j)}$ represents the average channel power gain, and $\eta_{\mathbf{d},m}^{(j)} \sim \mathrm{exp}(1)$ denotes the fading power gain. Additionally, $\beta =(4 \pi f_{c}/c)^{-2}$ is the average channel power gain at a reference distance 1m, where $f_{c} $ is the carrier frequency, and $c = 3.0 \times 10^{8}$ $(\mathrm{m/s})$ is the light speed.\\
\indent The cascaded BS-IRS-UE $0$ channel can be decomposed into  three components \cite{ref10}: BS-IRS transmission, IRS reflecting  and IRS-UE $0$ transmission, expressed as
\begin{equation}
	f_{\mathbf{ir},m}^{(j,q)} \triangleq [\mathbf{f}_{\mathbf{i},m}^{(j,q)}]^{T} \mathbf{\Psi}^{(q)} [\mathbf{f}_{\mathbf{r}}^{(q)}]^{T} = \sum_{n=1}^{N} f_{\mathbf{i},m,n}^{(j,q)} f_{\mathbf{r},n}^{(q)} e^{\mathbf{i} \psi_{n}^{(q)}} m \in  \Phi_{j}. \label{formula2}
\end{equation}
where $\mathbf{f}_{\mathbf{i},m}^{(j,q)}\triangleq[f_{\mathbf{i},m,1}^{(j,q)} , \dots , f_{\mathbf{i},m,N}^{(j,q)}]^{T}$ and $\mathbf{f}_{\mathbf{r}}^{(q)}\triangleq[f_{\mathbf{r},1}^{(q)}, \dots , f_{\mathbf{r},N}^{(q)}] $. 
The channel power gains from BS $m \in \Phi_{j}$ to the $n$-th reflecting element of IRS $q$, and from the $n$-th element to the \emph{typical} UE $0$ are, respectively, given by $|f_{\mathbf{i},m,n}^{(j,q)}|^{2}  \triangleq \eta_{\mathbf{i},m,n}^{(j,q)} l_{\mathbf{i},m}^{(j,q)}$ and $|f_{\mathbf{r},n}^{(q)}|^{2} \triangleq \eta_{\mathbf{r},n}^{(q)} l_{\mathbf{r}}^{(q)}$ where $l_{\mathbf{i},m}^{(j,q)} \triangleq \beta (r_{j,m,q}^{2} + (H_{j}^{2}-H_{I}^{2})) ^{-\alpha_{j}/2}$  and $l_{\mathbf{r}}^{(q)} \triangleq \beta (d_{q}^2 + H_{I}^2)^{-\alpha_{I}/2}$ denote the  corresponding average channel power gain.
\subsection{Performance Metrics} \label{subsec 2-3}
In this work, we aim to evaluate  the coverage probability and network  spatial throughput of the IRS-assisted HCN. The coverage probability is defined as the sum of the product of the association  probability and the conditional coverage probability $\mathbb{P}_{\mathrm{Cov}}^{k}|_{Z_{k},d_{0}}$ of each tier, which is given by
\begin{equation}
	\mathbb{P}_{\mathrm{Cov}}= \sum_{k=1}^{K} \mathcal{A}_{k}  \mathbb{E}_{Z_{k},d_{0}} [\mathbb{P}_{\mathrm{Cov}}^{k}|_{Z_{k},d_{0}} ],
\end{equation}
where  $\mathbb{P}_{\mathrm{Cov}}^{k}|_{Z_{k},d_{0}} =\mathbb{P}[\gamma_{k} > \gamma_{0}]|_{Z_{k},d_{0}}$ with $\gamma_{0}$  denoting the Signal-to-Interference-plus-Noise-Ratio (SINR) threshold and $\gamma_{k}=S_{k}/(I+\sigma^{2})$ being the conditional SINR when the \emph{typical} UE $0$ is associated with the $k$-th tier. Moreover, $S_{k}$ is the conditional desired signal power given by
\begin{equation}
	S_{k} \triangleq  P_{k} \cdot |f_{\mathbf{d},0}^{(k)} + \sum_{q \in \varDelta} f_{\mathbf{ir},0}^{(k,q)}|^{2},
	\label{formula18}
\end{equation}
and $I$ is the aggregated  received interference power given by
\begin{equation}
	I \triangleq \sum_{j=1}^{K}\sum_{m \in \Phi_{j} \backslash BS_{0}^{(k)}} P_{j} \cdot |f_{\mathbf{d},m}^{(j)} + \sum_{q \in \varDelta} f_{\mathbf{ir},m}^{(j,q)}|^{2},
	\label{formula19}
\end{equation}
where $ BS_{0}^{(k)}$  represents the $k$-th tier BS $0$. We denote the required achievable rate (in bps/Hz) of UE $0$ by $R_{0} \triangleq \mathrm{log}_{2}(1+\gamma_{0})$,
and the conditional spatial throughput can be expressed as $T_{k}=\mathbb{P}_{\mathrm{Cov}}^{k}R_{0}\lambda_{k}^{'}$. With the law of total probability, the network spatial throughput is denoted by 
\begin{equation}
	T \triangleq \sum_{k=1}^{K}\mathcal{A}_{k} T_{k}  =\sum_{k=1}^{K} \mathcal{A}_{k} \mathbb{P}_{\mathrm{Cov}}^{k} R_{0} p_{k} \lambda_{k}.
	\label{formula22}
\end{equation} 
\section{Performance Analysis}
\subsection{Channel  Statistics} \label{CPS}
\indent To facilitate the performance analysis, we need to first derive the channel power statistics caused by the involvement of IRSs.  Depending on whether IRS is associated to the UE $0$, the reflection channel can be divided into beamforming channel and random scattering channel. For the associated IRS $0$,  we assume perfect channel estimation such that the reflect signals from $N$ elements are the same phase as the direct signal at UE $0$. Therefore, we can obtain the amplitude of the cascaded BS $0$-IRS $0$-UE $0$ as
$|f_{\mathbf{ir},0}^{(k,0)}|=|\mathbf{f}_{\mathbf{i},0}^{(k,0)}|^{T}|\mathbf{f}_{\mathbf{r}}^{(0)}|=\sum_{n=1}^{N}|f_{\mathbf{ir},0,n}^{(k,0)}|$
 where each channel amplitude $|f_{\mathbf{ir},0,n}^{(k,0)}|$ is a double-Rayleigh random variable (RV) with independent $|f_{\mathbf{i},0,n}^{(k,0)}| \sim  \mathcal{R}(\sqrt{l_{\mathbf{i},0}^{(k,0)}/2})$ and $|f_{\mathbf{r},n}^{(0)}| \sim \mathcal{R}(\sqrt{l_{\mathbf{r}}^{(0)}/2})$ because of Rayleigh fading.  Therefore, we can derive the mean and variance of $|f_{\mathbf{ir},0,n}^{(k,0)}|$ as $\mathbb{E}\{ |f_{\mathbf{ir},0,n}^{(k,0)} | \}  \triangleq  \frac{\pi}{4}\sqrt{l_{\mathbf{i},0}^{(k,0)}l_{\mathbf{r}}^{(0)}} $ and $\mathrm{var}\{|f_{\mathbf{ir},b,n}^{(k,t)}|\} \triangleq (1-\frac{\pi^{2}}{16}l_{\mathbf{i},0}^{(k,0)}l_{\mathbf{r}}^{(0)})$, respectively.\\
 \indent Based on  the central limit theorem (CLT), the channel amplitude is approximated to follow the Gaussian distribution \textcolor{black}{for large $N$,} given by
 \begin{equation}
 	|f_{\mathbf{ir},0}^{(k,0)}| \stackrel{approx}{\sim}
 	\mathcal{N}(N \dfrac{\pi}{4}\sqrt{l_{\mathbf{i},0}^{(k,0)}l_{\mathbf{r}}^{(0)}}, N(1-\dfrac{\pi^{2}}{16})l_{\mathbf{i},0}^{(k,0)}l_{\mathbf{r}}^{(0)}).
 	\label{formula7}
 \end{equation}
\indent Consequently, for the cascaded BS $0$-IRS $0$-UE $0$ channel, the average signal power is equivalent to the second moment of $|f_{\mathbf{ir},0}^{(k,0)}|$ which is given by
\begin{equation}
	\mathbb{E}\{|f_{\mathbf{ir},0}^{(k,0)}|^{2}\} = [\dfrac{\pi^{2}}{16}N^{2}+(1-\dfrac{\pi^{2}}{16})N] \cdot l_{\mathbf{i},0}^{(k,0)}l_{\mathbf{r}}^{(0)} \triangleq G_{\mathrm{bf}} \cdot l_{\mathbf{i},0}^{(k,0)}l_{\mathbf{r}}^{(0)},
	\label{formula8}
\end{equation}
where $G_{\mathrm{bf}} \triangleq [\dfrac{\pi^{2}}{16}N^{2}+(1-\dfrac{\pi^{2}}{16})N]$ denotes the beamforming gain coefficient.  It shows that by performing reflect beamforming, IRS achieves a significant gain in the average signal power, which enhances with the growing reflecting element number $N$ in $O(N^2)$.\\
\indent From another aspect, for any IRS $q$ belongs to set $\varDelta$ except IRS $0$,  it just scatters  the incident signal from BSs, leading to a  uniformly random phase shift. According to the CLT, \textcolor{black}{for a practically large $N$,} we can approximate the combined scattering channel to the following CSCG distribution
\begin{equation}
	f_{\mathbf{ir},m}^{(j,q)} \stackrel{approx.}{\sim} \mathcal{CN}(0, Nl_{\mathbf{i},m}^{(j,q)}l_{\mathbf{r}}^{(q)}).
	\label{formula9}
\end{equation}
\indent Similar to  (\ref{formula8}), the average signal power of the cascaded BS $m$-IRS $q$-UE $0$ channel $f_{\mathbf{ir},m}^{(j,q)}$ is  derived as
\begin{equation}
	\mathbb{E}\{|f_{\mathbf{ir},m}^{(j,q)}|^{2}\} = N\cdot l_{\mathbf{i},m}^{(j,q)}l_{\mathbf{r}}^{(q)} \triangleq G_{sc} \cdot l_{\mathbf{i},m}^{(j,q)}l_{\mathbf{r}}^{(q)},
	\label{formula10}
\end{equation}
where  $G_{sc} \triangleq N$ denotes the random scattering gain coefficient of average channel power product $l_{\mathbf{i},m}^{(j,q)}l_{\mathbf{r}}^{(q)}$, which boosts with the growing number of reflecting elements $N$ in $O(N)$.
\subsection{Signal Power Distribution}\label{SPD}
\indent The signal power distribution depends on whether the  UE $0$ has an IRS $0 \in \varDelta$ to associate. Specifically, we need to discuss the following two cases: IRS reflect beamforming and IRS random scattering. 
\subsubsection{IRS Reflect Beamforming} In this case,  the associated IRS $0 \in \varDelta$  provides reflect beamforming to UE $0$ while the other IRSs in  $\varDelta$  performs the random scattering.  According to (\ref{formula18}), the overall conditional desired signal  is the summation of a Rayleigh distributed RV and $N$ Gaussion RVs which is difficult to obtain the exact PDF.  Since it is of great difficulty to derive the exact distribution of the conditional desired signal power,  
we  approximate $S_{k}$ as a Gamma distribution RV, i.e., $S_{k}|_{Z_{k},d_{0}} \sim \Gamma[\tau_{1},\theta_{1}]$ with $\tau_{1}$ and $\theta_{1}$ being the shape parameter and scale parameter, respectively. With the moment matching technique \cite{SamanAtapattu2020ReconfigurableIS}, we have
\begin{align}
	\tau_{1} \triangleq \frac{(\mathbb{E}\{S_{k}\}|_{Z_{k} , d_{0}})^{2}}{ \mathrm{var}\{S_{k}\}|_{Z_{k} , d_{0}}}, \quad 
	\theta_{1} \triangleq \frac{\mathrm{var}\{S_{k}\}|_{Z_{k} , d_{0}} }{\mathbb{E}\{S_{k}\}|_{Z_{k} , d_{0}}},
\end{align}
where the variance is given by  $\mathrm{var}\{S_{k}\}|_{Z_{k}}\triangleq \mathbb{E}\{S_{k}^{2}\}|_{Z_{k}, d_{0}} - (\mathbb{E}\{S_{k}\}|_{Z_{k} , d_{0}})^{2}$. To obtain  $\tau_{1}$ and $\theta_{1}$, we need to first derive the first and second moments of $S_{k}|_{Z_{k},d_{0}}$ conditioned on the locations of BSs and IRSs, which is given by Lemma \ref{lemma1}.
\newtheorem{lemma}{Lemma}
\begin{lemma} \label{lemma1}
	The first and second moments of the conditional desired signal power $S_{k}|_{Z_{k},d_{0}}$ are given by
	\begin{align}
		\mathbb{E}\{S_{k}\}|_{Z_{k} , d_{0}} &=P_{k} \cdot(\mathbb{E}\{|f_{1}|^{2}\}|_{Z_{k} , d_{0}} + \mathbb{E}\{|f_{2}|^{2}\}|_{Z_{k} , d_{0}} ),	  \label{32}\\
		\mathbb{E}\{S_{k}^{2}\}|_{Z_{k} , d_{0}} &=P_{k}^{2} \cdot (\mathbb{E}\{|f_{1}|^{4}\}|_{Z_{k} , d_{0}} + \mathbb{E}\{|f_{2}|^{4}\}|_{Z_{k} , d_{0}}  \notag \\
		&+ 4\mathbb{E}\{|f_{1}|^{2}\}|_{Z_{k} , d_{0}}\mathbb{E}\{|f_{2}|^{2}\}|_{Z_{k} , d_{0}}).
		\label{formula27}
	\end{align} 
where $f_{1}\triangleq f_{\mathbf{d},0}^{(k)} + f_{\mathbf{ir},0}^{(k,0)}$ and $f_{2}  \triangleq \sum_{q \in \varDelta \backslash \{0\}} f_{\mathbf{ir},0}^{(k,q)}$, respectively, represent the composite signals of a combination of the direct signal and the reflect beamforming signal with the same phase, and that of the summation of random scattering signal via IRSs $\in \varDelta\setminus \{0\}$. \\
Proof. See Appendix A for the detailed expressions of the first and second moments of $f_{1}$ and $f_{2}$.  $ \quad \quad \quad \quad \quad  \quad \quad \quad  \quad \quad \qedsymbol$ 
\end{lemma}
\subsubsection{Without IRS Assisted} For the case $\varDelta = \emptyset$, UE $0$ has no IRS to associate with.  Under the condition of being associated with the $k$-th tier, the signal power  $S_{k}|_{Z_{k},d_{0}}=|f_{\mathbf{d},0}^{(k)}|^{2}$ follows the Gamma distribution with the shape parameter being 1 which is equivalent to the exponential distribution with  $l_{\mathbf{d},0}^{(k)}$ being the mean value. In other words, we can define $S_{k}|_{Z_{k},d_{0}} \sim \Gamma[\tau_{2},\theta_{2}]$ where  the shape parameter $\tau_{2}=1$ and the scale parameter $\theta_{2}=l_{\mathbf{d},0}^{(k)}$. \\
\indent By combining the above two cases,  we can approximate the signal power conditioned on the associated $k$-th tier, BS $0$-UE $0$ distance  $Z_{k}$ and IRS $0$-UE $0$ distance $d_{0}$ as 
\begin{align}
	\begin{split}
		S_{k}|_{Z_{k},d_{0}} \stackrel{approx}{\sim} \Gamma[\tau,\theta]=	\left \{
		\begin{array}{ll}
			\Gamma[\tau_{1},\theta_{1}],   & \varDelta \neq \emptyset; \\
			\Gamma[\tau_{2},\theta_{2}],   & \varDelta = \emptyset. 
		\end{array}
		\right.
	\end{split}
	\label{sigo1}
\end{align}
\subsection{Interference Power Distribution}
 For a random BS $m$ in the $j$-th tier, both the direct interference channel and the cascaded BS $m$-IRS $q$-UE $0$ channel follow the CSCG distribution, and the composite interference channel $f_{\mathbf{d},m}^{(j)} + \sum_{q \in \varDelta} f_{\mathbf{ir},m}^{(j,q)}$ from the $j$-th tier BS $m$ follows the CSCG distribution with zero mean. As such, the composite interference power $I_{m}^{(j)} \triangleq P_{j}|f_{\mathbf{d},m}^{(j)} + \sum_{q \in \varDelta} f_{\mathbf{ir},m}^{(j,q)}|^{2}$ follows an exponential distribution, given by 
 \begin{equation}
 	I_{m}^{(j)} \triangleq \bar{I}_{m}^{(j)}\eta_{m}^{(j)}=P_{j}\biggl(l_{\mathbf{d},m}^{(j)}+N\sum_{q\in\varDelta}l_{\mathbf{i},m}^{(j,q)}l_{\mathbf{r}}^{(q)} \biggr)\eta_{m}^{(j)},
 	\label{imj}
 \end{equation}
where $\bar{I}_{m}^{(j)}$ is the average interference power and $\eta_{m}^{(j)} \stackrel{dist.}{=}\eta \sim \mathrm{exp}(1)$.  
Since we only consider the IRSs within the local region,  to simplify the analysis, we made a reasonable approximation that the 3D distance from a given BS to the \emph{typical} UE $0$ is equivalent to that from this BS to  IRSs $\in \varDelta$. Thus, we have $l_{\mathbf{d},m}^{(j)} \approx l_{\mathbf{i},m}^{(j,q)}$. By substituting (\ref{imj}) into (\ref{formula19}),  the aggregated interference can be expressed as
 \begin{equation}
	I \approx \rho \sum_{j=1}^{K} \sum_{m \in \Phi_{j}^{'} \setminus BS_{0}^{(k)}} P_{j}l_{\mathbf{d},m}^{(j)} \eta_{m}^{(j)}.
\end{equation} 
where $\rho \triangleq 1+\sum_{q \in \varDelta} l_{\mathbf{r}}^{(q)}$ denotes the relative  power gain of scattering paths with regards to  all IRSs  in $ \varDelta$.
\\
\indent In order to obtain the CDF of the conditional interference power, we need to derive the Laplace transform of interference. To simplify the expression of Laplace transform, we approximate $\rho$ by its mean value $\bar{\rho}$ instead. For the case $\varDelta \neq \emptyset$, the main scattering interference power comes from the nearest IRS $0$,
 $\bar{\rho} \triangleq 1+N l_{\mathbf{r}}^{(0)} +NE_{\mathrm{r1}}(d_{0})\triangleq K_{\mathrm{sc}}(d_{0})$. For the case $\varDelta = \emptyset$, there is no IRS scattering interference and we have $\bar{\rho}=1$.
  Thus, the Laplace transform of  $I|_{Z_{k},d_{0}}$ is given by
  \begin{equation}
  	\small
  	\begin{aligned}
  		&\mathcal{L}_{I}|_{Z_{k},d_{0}}(s)  \triangleq \mathbb{E}\{e^{-sI}\}|_{Z_{k},d_{0}} \notag \\
  		&\stackrel{(a)}{=} \prod_{j=1}^{K}\mathbb{E}_{\Phi_{j}^{'}} \Big\{\prod_{m \in \Phi_{j}^{'}\setminus BS_{0}^{(k)}}\mathbb{E}_{\eta}\{\mathrm{exp}(-s\bar{\rho}P_{j} l_{\mathbf{d},m}^{(j)} \eta ) \}   \Big\}|_{Z_{k},d_{0}} \notag \\
  		&\stackrel{(b)}{=} \prod_{j=1}^{K}\mathrm{exp}\Big(-2\pi\lambda_{j}^{'}\int_{X_{j,m}}^{\infty} \Big(1-\dfrac{1}{1+s\bar{\rho}P_{j} l_{\mathbf{d},m}^{(j)}}\Big) x\mathrm{d}x          \Big) \notag \\
  		&\stackrel{(c)}{=} \mathrm{exp}\Big(-2\pi\sum_{j=1}^{K} \lambda_{j}^{'}\int_{z_{j}}^{\infty}\Big(1- \dfrac{1}{1+(s\bar{\rho}P_{j}) \beta z^{-\alpha_{j}}}\Big)z  \mathrm{d}z\Big)\notag \\
  		&\stackrel{(d)}{=}\mathrm{exp}\Big(-2\pi\sum_{j=1}^{K}\lambda_{j}^{'}\mathrm{U}(s\bar{\rho}P_{j})\Big),
  		\label{formula apendx}
  	\end{aligned}
  \end{equation}
where (a) is follows from  the independence of $\Phi_{j}^{'}$ and (b) is due to the fact that   $\eta \sim \mathrm{exp}(1)$. The closest interference BS in the $j$-th tier at a horizontal distance $X_{j,m} =\sqrt{\mathrm{max}(0, z_{j}^2-H_{j}^2)+H_{j}^{2}}$ with $z_{j} \triangleq (\hat{P_{j}}\hat{B_{j}})^{1/\alpha_{j}}z_{k}^{1/\hat{\alpha}_{j}}$. In (c), we have $z \triangleq \sqrt{r^{2} +H_{j}^{2}}$  and $z_{k} \triangleq \sqrt{r_{k}^{2} +H_{k}^{2}} $. Finally, in (d), we rewrite the integral using the Gauss hypergeometric function, and define a function $\mathrm{U}(\cdot)$  as  
\begin{equation}
	\small
	\begin{aligned}
		\mathrm{U}(x) &\triangleq \frac{\pi}{\alpha_{j}\mathrm{sin}(\frac{2\pi}{\alpha_{j}})}(\beta x)^{\frac{2}{\alpha_{j}}} - \frac{z_{j}^{2}}{2} \cdot _2F_{1}\Big(1,\frac{2}{\alpha_{j}},1+\frac{2}{\alpha_{j}},-\frac{1}{\beta z_{j}^{-\alpha_{j}}x}\Big),
		\label{uuu}
	\end{aligned} 
\end{equation}
where $_2F_1$ is the Gauss  hypergeometric function. \\
\indent Based on (\ref{formula apendx}), we can get the CDF of the conditional interference power by taking the inverse Laplace transform as $F_{I|_{Z_{k},d_{0}}} =\mathcal{L}^{-1}[\frac{1}{s}\mathcal{L}_{I|_{Z_{k},d_{0}}}(s)]$
which can be computed  with the function packages
provided by Mathematica. 
\subsection{Coverage Probability and Spatial Throughput}
\indent According to Section \ref{subsec 2-3}, the conditional coverage probability  $\mathbb{P}_{\mathrm{Cov}}^{k}|_{Z_{k},d_{0}} \triangleq \mathbb{P}\{S_{k} > \gamma_{0}(I+\sigma^{2})\}|_{Z_{k},d_{0}}$. Since we have approximated the conditional signal power to follow the Gamma distribution, the conditional coverage probability is
\begin{align}
	\mathbb{P}_{\mathrm{Cov}}^{k}|_{Z_{k},d_{0}} &\approx \mathbb{E}_{I} \Big \{  \dfrac{1}{\Gamma(\tau)} \int_{\frac{\gamma_{0}(I+\sigma^{2})}{\theta}}^{\infty} t^{(\tau-1)} e^{-t} \mathrm{d}t   \Big\}\Big|_{Z_{k},d_{0}}.
	\label{formula48}
\end{align}
\indent For integer $\tau$, we have $\Gamma(\tau)= (\tau-1)!$, and using the approach of integration by parts, we can transform (\ref{formula48}) into  
\begin{align}
	\mathbb{P}_{\mathrm{Cov}}^{k}|_{Z_{k},d_{0}} \approx \sum_{i=0}^{\tau-1} \dfrac{(-1)^{i}}{i !} \dfrac{\partial^{i} }{\partial s^{i}} \Big[\mathbb{E}\{e^{-s\varpi}\}|_{Z_{k},d_{0}}\Big]_{s=1},
	\label{formula50}
\end{align}
where $\varpi \triangleq \gamma_{0}(I+\sigma^{2})/{\theta}$, and the detailed derivation is a modification of the Appendix D in \cite{ref14}.  Based on (\ref{formula apendx}), $\mathbb{E}\{e^{-s\varpi}\}|_{Z_{k},d_{0}}$  can be expressed as the Laplace transform $\mathcal{L}_{\varpi|_{Z_{k},d_{0}}}(s)$, given by
\begin{align}
	\mathcal{L}_{\varpi|_{Z_{k},d_{0}}}(s)
	&=\mathrm{exp}\biggl(-\dfrac{s\gamma_{0}\sigma^{2}}{\theta} - 2\pi \sum_{j=1}^{K} \lambda_{j}' \mathrm{U}(\frac{s\gamma_{0}\bar{\rho}P_{j}}{\theta})\biggr) \notag \\
	&\triangleq \mathrm{exp} \big(V(s)\big).
	\label{formula51}
\end{align}
\indent The conditional coverage can be calculated by deriving the ($\tau-1$)-order derivatives with regards to  $\mathrm{exp}(V(s))$ which can be derived by utilizing  $\mathrm{Fa\grave{a}}$ $\mathrm{di}$ $\mathrm{Bruno's}$ formula \cite{7544562}. \\  
\indent On the other hand, for  non-integer $\tau$, its upper and lower bounds can be used to approximate the conditional coverage probability.
\begin{equation}
	\mathbb{P}_{\mathrm{Cov}}^{k}|_{Z_{k},d_{0}} \approx \omega \mathbb{P}_{\mathrm{Cov}}^{k}|_{Z_{k},d_{0}, \lfloor \tau\rfloor} + (1-\omega)\mathbb{P}_{\mathrm{Cov}}^{k}|_{Z_{k},d_{0}, \lceil \tau\rceil},
	\label{formula52}
\end{equation}
where $\lceil \cdot \rceil$ and $\lfloor \cdot \rfloor$ are the ceiling and floor functions, \textcolor{black}{and the weight $\omega \triangleq M(\lceil \tau\rceil-\tau)/(M(\lceil \tau\rceil -\tau)+(\tau-\lfloor\tau\rfloor))$,}
where $M$ denotes  the priority factor to illustrate the non-linearity of $\mathbb{P}_{\mathrm{Cov}}^{k}|_{Z_{k},d_{0}}$ with $\tau$.\\
\indent Note that for the case $\varDelta \neq \emptyset$, it is too complicated to calculate for a large value of $\tau$. In fact, the value of $\mathbb{E}\{S_{k}\}|_{Z_{k},d_{0}} \gg \mathrm{var}\{S_{k}\}|_{Z_{k},d_{0}}$ when $\tau$ is large, and the signal power can be approximated as its mean value. Hence, when  $\tau$ is greater than the threshold, we can use the CDF of interference power to calculate  $\mathbb{P}_{\mathrm{Cov}}^{k}|_{Z_{k},d_{0}}$, given by
\begin{equation}
	\small
	\begin{aligned}
		\mathbb{P}_{\mathrm{Cov}}^{k}|_{Z_{k},d_{0}}
		&\approx \mathbb{P}\{I<\frac{\mathbb{E}\{S_{k}\}|_{Z_{k},d_{0}}}{\gamma_{0}} -\sigma^{2}\}|_{Z_{k},d_{0}} \triangleq F_{I|_{Z_{k},d_{0}}} (y),
		\label{666}
	\end{aligned}
\end{equation}
where $y\triangleq \mathbb{E}\{S_{k}\}|_{Z_{k},d_{0}}/\gamma_{0}-\sigma^{2} $. By averaging over $Z_{k}$ and $d_{0}$, we obtain the per-tier  coverage probability $\mathbb{P}_{\mathrm{Cov}}^{k}$. Finally, based on $\mathbb{P}_{\mathrm{Cov}}^{k}$ and association probability $\mathcal{A}_{k}$, we derive the unconditional overall coverage probability as
\begin{equation}
	\small
	\begin{aligned}
		\mathbb{P}_{\mathrm{Cov}} &\approx \sum_{k=1}^{K} \mathcal{A}_{k}  \biggl(\int_{z=H_{k}}^{\infty} \int_{d_{0}}^{D_{\mathrm{max}}} \mathbb{P}_{\mathrm{Cov}}^{k}|_{Z_{k},d_{0}} f_{d_{0}}(d_{0})f_{X_{k,0}}(z)\mathrm{d}d_{0}\mathrm{d}z  \\ 
		&+ \mathbb{P}[\varDelta =\emptyset] \int_{z=H_{k}}^{\infty} \mathbb{P}_{\mathrm{Cov}}^{k}|_{Z_{k},d_{0}>D_{\mathrm{max}}} f_{X_{k,0}}(z)\mathrm{d}z\biggr).
		\label{53}
	\end{aligned}	
\end{equation}
where $\mathbb{P}[\varDelta = \emptyset] \triangleq e^{-\lambda_{I}\pi D_{\mathrm{max}}^{2}}$  , and  $\mathbb{P}_{\mathrm{Cov}}^{k}|_{Z_{k},d_{0}>D_{\mathrm{max}}}$ represents the conditional coverage probability for the case  $\varDelta = \emptyset$.
By substituting (\ref{53}) into (\ref{formula22}), we can obtain the network spatial throughput.
\section{Numerical Results}
\begin{table}[t]
	\begin{center} 
		\caption{\\ NOTATION AND DEFAULT VALUES.}
		\label{tab1}
		\begin{tabular}{| c | c | c |}
			\hline
			Notation & Description & Default Value\\
			\hline
			$P_{1}, P_{2}$ & Macrocell, Picocell  Transmit power & 53 $\mathrm{dBm}$, 33 $\mathrm{dBm}$\\
			\hline
			$H_{1}, H_{2}$ & Macrocell, Picocell BS height  &
			20 $\mathrm{m}$, 10 $\mathrm{m}$\\
			\hline
			$\lambda_{1}, \lambda_{2}$ & Macrocell, Picocell BS density  &
			10 $\lambda_{0}$,  50 $\lambda_{0}$\\
			\hline
			$\alpha_{1}, \alpha_{2}$& Macrocell, Picocell path loss exponent &4, 3.5\\
			\hline
			$H_{I}$& IRS height & 1 $\mathrm{m}$\\
			\hline
			$N$&  Number of IRS reflecting elements &  1000\\
			\hline
			$\lambda_{I}$ & IRS density &  200 $\lambda_{0}$\\
			\hline
			$\alpha_{I}$ & IRS path loss exponent & 3\\
			\hline
			$f_{c}$& Carrier frequency & 2 $\mathrm{GHz}$\\
			\hline
			$M$ & priority factor & 0.6 \\
			\hline
			$R_{0}$& Achievable rate threshold & 1  $\mathrm{bps/Hz}$\\
			\hline
			$D_{\mathrm{max}}$& IRS local region radius & 50 $\mathrm{m}$\\
			\hline
			$\sigma^{2}$& Additive White Gaussian Noise  & -117 $\mathrm{dBm}$\\
			\hline
		\end{tabular}
	\end{center}
\end{table}
\begin{figure*}
\vspace{-1em}
		\begin{minipage}{0.33\textwidth}
			\includegraphics[width=1\textwidth]{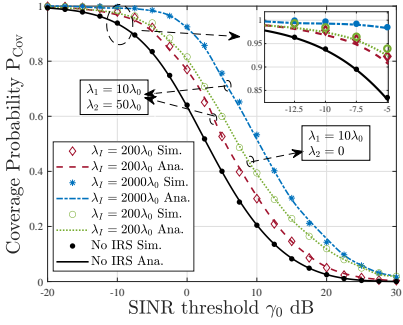}
			\captionsetup{font={scriptsize}}
			\caption{ Coverage probability as a function of SINR threshold for different picocell BS density $\lambda_{2}$ and IRS density $\lambda_{I}$.}
			\label{fi2}
		\end{minipage}
		\begin{minipage}{0.33\textwidth}
			\includegraphics[width=1\textwidth]{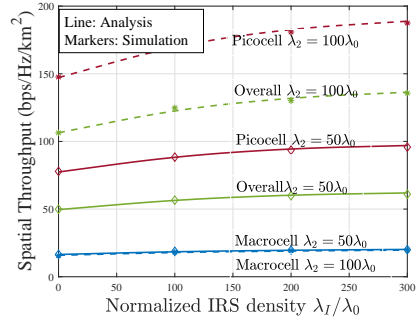}
			\captionsetup{font={scriptsize}}
			\caption{Spatial throughput as a function of normalized IRS density $\lambda_{I}$ for different picocell BS densities $\lambda_2$.}
			\label{fi4}
		\end{minipage}
		\begin{minipage}{0.33\textwidth}
			\includegraphics[width=1\textwidth]{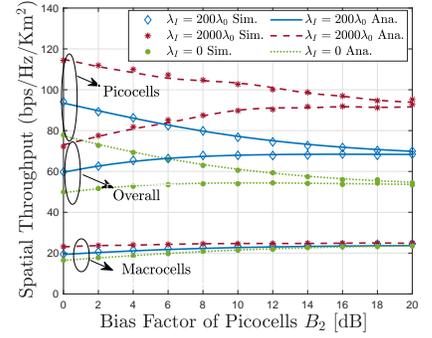}
			\captionsetup{font={scriptsize}}
			\caption{Spatial throughput as a function of picocell BS bias factor $B_{2}$ for different normalized IRS density $\lambda_{I}$}
			\label{fi5}
		\end{minipage}
	\vspace{-1em}
\end{figure*}
In this section, we verify the analytical  results via simulations by considering an IRS assisted two-tier HCN scenario  consists of a first tier of macro cell BSs overlaid with a second tier of pico cell BSs.  We consider a square window of  $ 4000 \times 4000$ $\mathrm{m}^{2}$ and denote $\lambda_{0} = 5 \times 10^{-6} /\mathrm{m^{2}}$ as the baseline density. Unless otherwise specified, we adopts the default values of the system parameters in Table~\ref{tab1}.\\
\indent In Fig. \ref{fi2}, we  evaluate the coverage probability of the IRS-assisted HCN.  We observe that the coverage enhancement achieved by IRS can be large than 0.3 in HCN, and a larger IRS density leads to a higher network coverage.
This can be explained by the fact that the gain of IRS passive beamforming grows in $O(N^{2})$ and $O(d_{0}^{-\alpha_{I}/2})$ while the interference contributed by random scattering grows in $O(N)$. Growing the IRS density decreases the distance $d_{0}$ between the \emph{typical} UE $0$ and the nearest IRS $0$, leading to an increase in the coverage. Compared with the IRS-assisted single tier network, the deployment of the picocell networks with lower path loss exponent can only degrade the overall coverage probability.\\
\indent In Fig. \ref{fi4}, we depict the spatial throughput as a function of normalized IRS density for different picocell BS densities $\lambda_2$.  We observe that the overall throughput is mainly contributed by the picocells due to the higher density of $\lambda_2$. It also shows that as IRS density $\lambda_I$ increases, the spatial throughput achieved by picocells first increases and then converges, while the spatial throughput achieved by macrocell nearly unchanged. This is because the density of picocells is much larger than that of macrocells, and thus, the typical link distance in picocell  is shorter than macrocell. Therefore, the beamforming gain achieved by IRSs is more obvious for the shorter distance between picocell BS and IRS.  \\ 
\indent In Fig. \ref{fi5}, we reveal the impact of the bias of picocells $B_{2}$ on spatial throughput with different IRS densities. It is observed that a growing $B_{2}$ decreases (increases) the spatial throughput achieved by picocells (macrocells), leading to the growth in overall network spatial throughput.  In fact, as  $B_{2}$ grows, a part of UEs of poor channel condition that are originally associated with macrocells will be forced to connect to the picocells, which enhances (lowers) the per-tier coverage probability of macrocells (picocells). Referring to (\ref{formula22}), the variation of overall spatial throughput depends on the product $\mathcal{A}_{k}\mathbb{P}_{\mathrm{Cov}}^{k} p_{k} \lambda_{k}$ in each tier.
We observe that the enhanced spatial throughput of macrocells dominates $T$, leading to the increase of $T$.  
Besides, as $B_{2}$ further rises, almost all UEs connect to the picocell network, and the overall spatial throughput converges to that achieved by picocell tier.  

\section{Conclusion}
In this work, we proposed an analytical framework to characterize the network performance of IRS-assisted $K$-tier HCN, in terms of coverage probability and network spatial throughput. We also characterized the average signal power and interference power in a randomly located BSs/IRSs scenario, and quantified the channel power gain achieved by IRSs. The analytical results are validated by extensive simulations, which shows that IRSs can significantly enhance the network performance of HCN while not changing the asymptotic behaviour of the conventional HCN.  


\appendices
\section{The First And Second Moments Of $f_{1}$ And $f_{2}$} \label{AP}
With the similar method proposed in Appendix B of \cite{ref14}, the first and secon d moments of $f_{1}$, $f_{2}$ are all zero because of the random phase. And the first and second moments of $|f_{1}|^{2}$ and $|f_{2}|^{2}$ can be derived by
\begin{align}	
	&\mathbb{E}\{|f_{1}|^{2}\}|_{Z_{k},d_{0}}=l_{\mathbf{d},0}^{(k)}\Big(1+N\frac{\pi}{4}\sqrt{\pi l_{\mathbf{r}}^{(0)}}+G_{\mathrm{bf}}l_{\mathbf{r}}^{(0)}\Big), \label{f44} \\
	&\mathbb{E}\{|f_{2}|^{2}\}|_{Z_{k},d_{0}}=Nl_{\mathbf{d},0}^{(k)}E_{\mathrm{r1}}(d_{0}),\label{f55} \\
	&\mathbb{E}\{|f_{2}|^{4}\}|_{Z_{k},d_{0}}=2N^{2}[l_{\mathbf{d},0}^{(k)}]^{2}E_{\mathrm{r3}}(d_{0}),
	\label{ffff} \\
	&\mathbb{E}\{|f_{1}|^{4}\}|_{Z_{k},d_{0}}=[l_{\mathbf{d},0}^{(k)}]^{2}\bigg[2+\frac{3}{4}\pi^{\frac{3}{2}}N\sqrt{l_{\mathbf{r}}^{(0)}}+6G_{\mathrm{bf}}l_{\mathbf{r}}^{(0)} \notag \\
	&+2\sqrt{\pi}\Big(\frac{\pi^{3}N^{3}}{64}+\frac{3\pi+ N^{2}(1-\frac{\pi^{2}}{16})}{4}\Big)[\l_{\mathbf{r}}^{(0)}]^{\frac{3}{2}}+  \notag \\
	&\Big(\frac{\pi^{4}N^{4}}{256}+\frac{3\pi^{2} N^{3}(1-\frac{\pi^{2}}{16})}{8}+3N^{2}(1-\frac{\pi^{2}}{16})^{2}\Big)[l_{\mathbf{r}}^{(0)}]^{2} \bigg], \label{f33} 
\end{align}
%
where 
$E_{\mathrm{r1}}(d_{0})$ denotes the expectation of $\sum_{q \in \varDelta \setminus \{0\}}l_{\mathbf{r}}^{(q)}$ which is given by
\begin{equation}
	\small
	\begin{aligned}
		E_{\mathrm{r1}}(d_{0})  =\dfrac{2\pi\lambda_{I}\beta}{\alpha_{I}-2}\Big[ (d_{0}^{2}+H_{I}^{2})^{\frac{2-\alpha_{I}}{2}}  - (D_{\mathrm{max}}^{2} +H_{I}^{2} )^{\frac{2-\alpha_{I}}{2}} \Big],
		\label{fff36}
	\end{aligned}
\end{equation}
and
\begin{equation}
	\small
\begin{aligned}
	E_{\mathrm{r3}} \triangleq \mathbb{E}\Big\{\Big(\sum_{q \in \varDelta \setminus \{0\}}l_{\mathbf{r}}^{(q)} \Big)^{2}  \Big\} =(E_{\mathrm{r1}}(d_{0}))^{2}+E_{\mathrm{r2}}(d_{0}),
\end{aligned}
\end{equation}
with $E_{\mathrm{r2}}(d_{0})\triangleq 
\frac{\pi \lambda_{I} \beta^{2}}{\alpha_{I}-1}[(d_{0}^{2}+H_{I}^{2})^{1-\alpha_{I}}-(D_{\mathrm{max}}^{2}+H_{I}^{2})^{1-\alpha_{I}}]$.\\


%
%
%
\footnotesize
\bibliographystyle{IEEEtran}
\bibliography{document}

\end{document}